# Canonical-ensemble extended Lagrangian Born–Oppenheimer molecular dynamics for the linear scaling density functional theory

Teruo Hirakawa [1], Teppei Suzuki [2,3], David R. Bowler [2,4,5] and Tsuyoshi Miyazaki [2]

[1] Department of Precision Science and Technology, Graduate school of Engineering, Osaka University, 2-1 Yamada-oka, Suita, Osaka 565-0871, Japan

[2] International Center for Materials Nanoarchitectonics, National Institute for Materials Science (NIMS), 1-1 Namiki, Tsukuba, Ibaraki 305-0044, Japan

[3] Computational Astrophysics Laboratory, RIKEN, 2-1 Hirosawa, Wako, Saitama 351-0198, Japan

[4] London Centre for Nanotechnology, University College London, 17-19 Gordon St, London, WC1H 0AH, United Kingdom

[5] Department of Physics and Astronomy, University College London, Gower St, London, WC1E 6BT, United Kingdom

E-mail: teppei.suzuki@riken.jp and tsuyoshi.miyazaki@nims.go.jp

(April 27, 2017)

## Abstract

We discuss the development and implementation of a constant temperature (NVT) molecular dynamics scheme that combines the Nosé–Hoover chain thermostat with the extended Lagrangian Born–Oppenheimer molecular dynamics (BOMD) scheme, using a linear scaling density functional theory (DFT) approach. An integration scheme for this canonical-ensemble extended Lagrangian BOMD is developed and discussed in the context of the Liouville operator formulation. Linear scaling DFT canonical-ensemble extended Lagrangian BOMD simulations are tested on bulk silicon and silicon carbide systems to evaluate our integration scheme. The results show that the conserved quantity remains stable with no systematic drift even in the presence of the thermostat.

## 1. Introduction

Quantum-mechanical molecular dynamics (MD) methods are based on accurate descriptions of interatomic forces calculated 'on the fly' from the electronic structure of the system. Density functional theory (DFT) is often the method of choice for electronic structure calculations because of its practical usefulness and affordability [1–3]. Currently, two major approaches for performing *ab initio* MD are the Born–Oppenheimer (BO) MD and the Car–Parrinello MD method [4, 5]. When performing BOMD, several extrapolation techniques [6–9] have been developed to predict the charge density and/or wavefunctions at each step; these methods improve the efficiency significantly, but break time-reversal symmetry, and can lead to a systematic energy drift. Recently, the extended Lagrangian BOMD formalism [10, 11] was developed to permit efficient propagation of the electronic degrees of freedom without breaking time-reversal symmetry, hence removing the systematic drift.

In the extended Lagrangian BOMD formalism, the auxiliary electronic degrees of freedom that is propagated by a time-reversible integrator is used as an efficient initial guess for self-consistent field (SCF) optimizations [10, 11]. Owing to the time-reversibility, the dynamics maintains the long-term stability of energy with a reduced number of SCF cycles (typically, only a few diagonalizations per time step are required) [12–16]. The extended Lagrangian BOMD scheme has been successfully applied to a variety of electronic structure methods, such as self-consistent tight-binding theory [13, 14] (also called density functional tight-binding theory), Hartree–Fock theory with Gaussian orbitals [15], and DFT within the plane-wave pseudopotential framework [16].

Importantly, the extended Lagrangian BOMD formalism is compatible not only with exact diagonalization but also with linear scaling approaches [17–20]. There has been a considerable recent interest in linear scaling electronic structure calculation methods [21–24]. This is because a major bottle neck of the traditional DFT method is the computational cost that scales cubically with respect to the number of the electrons in a system. In practice, systems consisting of $10^3$–$10^5$ atoms are currently beyond the scope of *ab initio* MD methods based on traditional approaches. Nonetheless, such system sizes are of particular relevance for materials science [25, 26], geophysics [27], and molecular biology [28]. Linear scaling electronic structure calculations on millions of atoms have been demonstrated [29, 30]; but calculations on systems with sizes of $10^3$–$10^5$ atoms are more practical targets for practical demonstrations of linear scaling electronic structure calculations.

The existing extended Lagrangian BOMD schemes have been extended to constant temperature studies by Martínez and coworkers [31]. They demonstrated that three different types of temperature-control methods—Langevin dynamics and the Andersen and Nosé thermostats—are compatible with the extended Lagrangian BOMD formalism with long-term energy conservation. Following this work, the Nosé–Hoover chain method [32–34], an extension of the Nosé dynamics, is expected to be compatible with the extended Lagrangian BOMD scheme. In the previous work [31], the integration scheme for Nosé dynamics was based on a simple leap-frog-like algorithm. In the case of Nosé–Hoover chain dynamics, however, time-reversible integrators based on the Liouville operator formulation have been widely used in MD simulations [35, 36]. In fact, the Liouville operator factorization technique has been applied to the Car–Parrinello MD method [37]; therefore, a similar decomposition scheme is expected to be appropriate for the extended Lagrangian BOMD formalism.

The purpose of the present work is twofold. One is to provide a more general framework for an integration scheme for canonical-ensemble (NVT) extended Lagrangian BOMD in the context of the Liouville operator formulation. The other is to give an integration scheme for linear scaling DFT canonical-ensemble simulations within the extended Lagrangian BOMD formalism, using the density matrix minimization (DMM) method [22] implemented in CONQUEST [38–41]. The rest of the paper is organized as follows. In Methodology, we briefly review extended Lagrangian formalism and the Nosé–Hoover chain method. We derive an integration scheme on the basis of the classical Liouville propagator for the extended Lagrangian BOMD scheme in the canonical ensemble. Finally, we describe a molecular dynamics scheme for the canonical-ensemble extended Lagrangian BOMD scheme in the framework of the DMM method. NVT extended Lagrangian BOMD simulations are tested on bulk silicon and silicon carbide systems to evaluate our integration scheme. In the end, we give our conclusions.

## 2. Methodology

*2.1. Extended Lagrangian Born–Oppenheimer molecular dynamics*

The key to the extended Lagrangian BOMD formalism proposed by Niklasson is to introduce the auxiliary electronic degrees of freedom that can be propagated via time-reversible integrators and used as a good initial guess for the SCF optimization [10, 11]. The extended Lagrangian is given by

$$\mathcal{L}^{\text{XLBO}} = \frac{1}{2}\sum_{I=1}^{N} M_I \dot{\mathbf{R}}_I^2 - E_{\text{tot}}[\{\mathbf{R}_I; D\}] + \frac{1}{2}\mu \text{Tr}[\dot{P}^2] - \frac{1}{2}\mu\omega^2 \text{Tr}[(D-P)^2],$$

(1)

where the potential energy $E_{\text{tot}}$ is defined by the self-consistent ground state $D$ for nuclear positions $\{\mathbf{R}_I\}$ with masses $\{\mathbf{M}_I\}$; this can take the form of the density matrix or the charge density for the system. The auxiliary electronic degrees of freedom $P$ evolves in a harmonic well centred on the SCF ground state solution $D$ and can be used as a good initial guess for the SCF optimization [10, 11]. The parameter $\mu$ is a fictitious electronic mass and $\omega$ is the frequency that determines the curvature of the harmonic well [11, 12].

In the limit of $\mu \to 0$, the Euler–Lagrange equations of motion derived from the extended Lagrangian (equation (1)) should read

$$M_I \ddot{\mathbf{R}}_I = -\frac{\partial E_{\text{tot}}[\{\mathbf{R}_I; D\}]}{\partial \mathbf{R}_I} = \mathbf{F}_I,$$

$$\ddot{P} = \omega^2 (D - P).$$

(2)

The first equation of motion is the normal BOMD equation for the ionic degrees of freedom, where the forces on the nuclei $\{\mathbf{F}_I\}$ are calculated at the self-consistent electronic ground state $D$. The second equation of motion is that for the auxiliary electronic variable $P$, which ensures the time-reversal symmetry in the propagation of the electronic degrees of freedom [10, 11]. The two equations of motion do not include the parameter $\mu$ [11]. The value of $\omega$ is associated with a given integration time step $\Delta t$.

*2. 2. Temperature control*

In Nosé–Hoover chain dynamics proposed by Martyna, Klein, and Tuckerman [34, 35], a physical system is coupled to a chain of imaginary thermostats. An extension to Nosé–Hoover chain dynamics in the framework of the extended Lagrangian BOMD formulation can be written as

$$M_I \ddot{\mathbf{R}}_I = -\frac{\partial E_{\text{tot}}}{\partial \mathbf{R}_I} - v_{\eta_1} M_I \dot{\mathbf{R}}_I,$$

$$\ddot{P} = \omega^2 (D - P),$$

$$Q_1\dot{\nu}_{\eta_1} = \left(\sum_I M_I \dot{\mathbf{R}}_I^2 - gkT\right) - \nu_{\eta_2} Q_1 \nu_{\eta_1},$$

$$Q_j\dot{\nu}_{\eta_j} = \left(Q_{j-1}\nu_{\eta_{j-1}}^2 - kT\right) - \nu_{\eta_{j+1}} Q_j \nu_{\eta_j},$$

$$Q_M\dot{\nu}_{\eta_M} = \left(Q_{M-1}\nu_{\eta_{M-1}}^2 - kT\right),$$

$$\dot{\eta}_j = \nu_{\eta_j}.$$

(3)

This formulation can be viewed as an extension of the previous work by Martínez *et al.* [31]. The variables $\{\eta_j\}$ and $\{\nu_{\eta_j}\}$ are dimensionless positions and velocities, respectively, for a chain of *M* thermostats. The parameter *g* is the number of the ionic degrees of freedom; *k* is the Boltzmann constant; *T* is the desired ionic temperature; and the parameters $\{Q_j\}$ are called fictitious thermostat masses that can be rewritten as $Q_1 = NkT/\omega_{\text{bath}}^2$ and $Q_j = kT/\omega_{\text{bath}}^2$. The value of the thermostat frequency $\omega_{\text{bath}}$ depends primarily on a characteristic phonon frequency of a system. In Nosé–Hoover chain dynamics, the first thermostat velocity $\nu_{\eta_1}$ acts as a dynamics friction coefficient [33]; it cools down or heats up the physical system if the instantaneous temperature becomes higher or lower than *T*. Notice that the auxiliary electronic degrees of freedom is indirectly coupled to the ionic fluctuations controlled by the thermostat.

In addition, the dynamics has the conserved quantity [34, 35]:

$$H' = \frac{1}{2}\sum_{I=1}^{N} M_I \dot{\mathbf{R}}_I^2 + E_{\text{tot}}[\{\mathbf{R}_I\}] + \frac{1}{2}\sum_{j=1}^{M} Q_j \nu_{\eta_j}^2 + gkT\eta_1 + \sum_{j=2}^{M} kT\eta_j.$$

(4)

The conserved quantity can be used to assess the quality of MD simulations [31, 35]. Since the auxiliary variable is used for a starting guess for the self-consistent optimization, the auxiliary variable *P* does not appear in the conserved quantity. For conventional BOMD, the quantity would not be well defined owing to the broken time-reversal symmetry [31].

*2. 3. Integration scheme*

On the basis of the Liouville operator formulation, we derive an integration scheme for Nosé–Hoover chain thermostatted extended Lagrangian BOMD. The Liouville operator for the equation of motion for such extended Lagrangian BOMD can be written as a sum of five parts:

$$iL = iL_1 + iL_2 + iL_3 + iL_4 + iL_{\text{NHC}},$$

(5)

which are defined as

$$iL_1 = \dot{P} \cdot \nabla_P,$$

$$iL_2 = \omega^2(D-P) \cdot \nabla_{\dot{P}},$$

$$iL_3 = \sum_{I=1}^{N} \dot{\mathbf{R}}_I \cdot \nabla_{\mathbf{R}_I},$$

$$iL_4 = \sum_{I=1}^{N} \left[\frac{\mathbf{F}_I}{M_I}\right] \cdot \nabla_{\dot{\mathbf{R}}_I},$$

(6)

with the operator $iL_{\text{NHC}}$ denoting the Nosé–Hoover chain part [35]:

$$iL_{\text{NHC}} = -\sum_{I=1}^{N} \nu_{\eta_1} R_I \cdot \nabla_{\dot{\mathbf{R}}_I} + \sum_{j=1}^{M} \nu_{\eta_j} \frac{\partial}{\partial \eta_j} + \sum_{j=1}^{M-1} \left(G_j - \nu_{\eta_j}\nu_{\eta_{j+1}}\right) \frac{\partial}{\partial \nu_{\eta_j}} + G_M \frac{\partial}{\partial \nu_M},$$

(7)

and

$$G_1 = \frac{1}{Q_1}\left(\sum_{I=1}^{N} M_I \dot{\mathbf{R}}_I^2 - gkT\right),$$

$$G_j = \frac{1}{Q_j}\left(Q_{j-1}\nu_{\eta_{j-1}}^2 - kT\right) \quad (j > 1).$$

(8)

In order to generate a time-reversible integrator, an approximate propagator for a time step $\Delta t$ is constructed:

$$\exp(iL\Delta t) = \exp\left(iL_{\text{NHC}}\frac{\Delta t}{2}\right)\exp\left(iL_4\frac{\Delta t}{2}\right)\exp\left(iL_3\frac{\Delta t}{2}\right)$$
$$\times \exp\left(iL_2\frac{\Delta t}{2}\right)\exp(iL_1\Delta t)\exp\left(iL_2\frac{\Delta t}{2}\right)$$
$$\times \exp\left(iL_3\frac{\Delta t}{2}\right)\exp\left(iL_4\frac{\Delta t}{2}\right)\exp\left(iL_{\text{NHC}}\frac{\Delta t}{2}\right) + \mathcal{O}(\Delta t^3),$$

(9)

where a symmetric Trotter expansion has been applied. Denoting the propagator for the Nosé–Hoover chain part as $U_{\text{NHC}}(\Delta t/2)$ and the propagator for the extended Lagrangian BOMD as $U_{\text{XL}}(\Delta t)$, we can rewrite the full propagator equation (9) as

$$U_{\text{XL-NVT}}(\Delta t) = U_{\text{NHC}}\left(\frac{\Delta t}{2}\right) U_{\text{XL}}(\Delta t) U_{\text{NHC}}\left(\frac{\Delta t}{2}\right) + \mathcal{O}(\Delta t^3).$$

(10)

Equations (9) and (10) are a symmetric decomposition that is accurate to $\mathcal{O}(\Delta t^3)$, which preserves the time-reversal symmetry. The propagator $e^{iL_2\Delta t/2}e^{iL_1\Delta t}e^{iL_2\Delta t/2}$ in $U_{\text{XL}}(\Delta t)$ is equivalent to the velocity Verlet algorithm for the auxiliary electronic variable $P$. Eliminating $\dot{P}$ in the algorithm yields the coordinate version of the Verlet integrator [36], which has been used in extended Lagrangian BOMD:

$$P(t+\Delta t) = 2P(t) - P(t-\Delta t) + \Delta t^2 \omega^2 \big(D(t) - P(t)\big). \tag{11}$$

Using the direct translation technique [35] and noting that the propagators $e^{iL_3\Delta t/2}e^{iL_4\Delta t/2}$ and $e^{iL_4\Delta t/2}e^{iL_3\Delta t/2}$ (applied before and after the propagator $e^{iL_2\Delta t/2}e^{iL_1\Delta t}e^{iL_2\Delta t/2}$) can form the velocity Verlet algorithm for the ionic degrees of freedom, we obtain a combined Verlet algorithm for the propagator $U_{\text{XL}}(\Delta t)$:

$$\begin{aligned}
\dot{\mathbf{R}}_I &\leftarrow \dot{\mathbf{R}}_I + \frac{\Delta t}{2M_I}\mathbf{F}_I, \\
\mathbf{R}_I &\leftarrow \mathbf{R}_I + \Delta t \dot{\mathbf{R}}_I, \\
P(t+\Delta t) &= 2P(t) - P(t-\Delta t) + \Delta t^2 \omega^2 \big(D(t) - P(t)\big), \\
\dot{\mathbf{R}}_I &\leftarrow \dot{\mathbf{R}}_I + \frac{\Delta t}{2M_I}\mathbf{F}_I.
\end{aligned} \tag{12}$$

The ground state $D(t+\Delta t)$ can be obtained using the auxiliary variable $P(t+\Delta t)$ as a starting guess for SCF optimization, and the force $\mathbf{F}_I(t+\Delta t)$ can be calculated using the ground-state potential that is determined by the atomic configurations. To propagate the Nosé–Hoover chain part $U_{\text{NHC}}(\Delta t/2)$, on the other hand, we can use a higher-order integration scheme:

$$U_{\text{NHC}}\left(\frac{\Delta t}{2}\right) = \prod_{j=1}^{n_{\text{ys}}} \exp\left(iL_{\text{NHC}} w_j \frac{\Delta t}{2}\right), \tag{13}$$

where $n_{\text{ys}}$ defines the order of the Yoshida–Suzuki integrator and $\{w_j\}$ are the integration parameters associated with the Yoshida–Suzuki algorithm [35]; a detailed algorithm of the propagator $U_{\text{NHC}}(\Delta t/2)$ has been proposed by Martyna *et al.* [35].

The first step of the integration scheme is to propagate the Nosé–Hoover chain thermostats for half a time step (part (a) in Algorithm I). The second step is to propagate the physical system via the operator $U_{\text{XL}}(\Delta t)$ (part (b) in Algorithm I). For the ionic degrees of freedom, the velocities of the atoms are propagated for half a time step and the positions are updated for the full time step. Subsequently, the auxiliary electronic degrees of freedom are propagated using a modified Verlet integrator:

$$P(t+\Delta t) = 2P(t) - P(t-\Delta t) + \kappa(D(t) - P(t)) + \alpha \sum_{k=0}^{K} c_k P(t - k\Delta t). \quad (14)$$

Here we have introduced a dissipation scheme to counteract the accumulation of numerical noise, which has been suggested by Niklasson *et al.* [12]. It has been shown that weak dissipation in the propagation of the auxiliary degrees of freedom is essential for the long-term stability of the trajectories [12, 13]. Detailed information about the dimensionless factor $\kappa = \Delta t^2 \omega^2$, the coupling parameter $\alpha$, the coefficients $\{c_k\}$, and the dissipation parameter $K$ has been reported by Niklasson *et al.* [12]. After the SCF optimization, the forces $\boldsymbol{F}_I(t+\Delta t)$ are computed from the ground-state potential surface and the velocities are propagated for the second half of the time step, thereby completing the velocity Verlet algorithm. Finally, the thermostat variables are propagated using the propagator $U_{\mathrm{NHC}}(\Delta t/2)$ for the second half of the time step (part (c) in Algorithm I).

*2. 4. Application to linear scaling DFT*

It was proposed by Arita, Bowler, and Miyzaki [18] that an extended Lagrangian BOMD scheme with the DMM method can be described as

$$\mathcal{L}^{\mathrm{XL-DMM}} = \frac{1}{2}\sum_{I=1}^{N} M_I \dot{\mathbf{R}}_I^2 + E_{\mathrm{tot}}[\{\mathbf{R}_I\}] + \frac{1}{2}\mu \mathrm{Tr}[\dot{X}^2] - \frac{1}{2}\mu\omega^2 \mathrm{Tr}[(LS - X)^2]. \quad (15)$$

Here the auxiliary variable $X$ evolves through a harmonic oscillator centred on the ground state characterized by the matrix $LS$, where the matrices $L$ and $S$ are the auxiliary density matrix and the overlap matrix between a set of localized orbitals (referred to as *support functions*), respectively. In the DMM method, weak idempotency requirement is imposed by expressing the density matrix $K$ in terms of $L$ and $S$ as

$$K = 3LSL - 2LSLSL. \quad (16)$$

Full details of the DMM method can be found in the literature [25, 35, 37]. In the limit of $\mu \to 0$, the Euler–Lagrange equations of motion derived from the extended Lagrangian should read

$$M_I \ddot{\mathbf{R}}_I = -\frac{\partial E_\text{tot}}{\partial \mathbf{R}_I},$$

$$\ddot{X} = \omega^2(LS - X).$$

(17)

An integration scheme for the canonical-ensemble extended Lagrangian BOMD scheme in the framework of the DMM method is summarized in Algorithm II. The integration scheme is essentially similar to that in Algorithm I, but is a little more complicated (see part (b) in Algorithm II). As is the case in Algorithm I, the first step of the integration scheme is to propagate the Nosé–Hoover chain thermostats for half a time step (part (a) in Algorithm II). In the second step (part (b) in Algorithm II), the velocities of the atoms are propagated for half a time step and the positions are updated for the full time step. Subsequently, the auxiliary variable $X$ can be integrated:

$$X(t + \Delta t) = 2X(t) - X(t - \Delta t) + \kappa\big(LS(t) - X(t)\big) + \alpha \sum_{k=0}^{K} c_k X(t - k\Delta t),$$

(18)

where a dissipation scheme is used. In this formulation, the matrix $XS^{-1}$ can be used as an initial guess for the DMM procedure:

$$L(t + \Delta t) = \text{DMM}[X(t + \Delta t)S^{-1}(t + \Delta t)].$$

(19)

In the present work, the matrix $S^{-1}$ is computed using Hotelling's method [42] (the matrix $S^{-1}$ has to be calculated at each time step). The forces on nuclei are computed for the ground state and the velocities are then propagated for the second half of the time step. The thermostat variables are then propagated using the propagator $U_\text{NHC}(\Delta t/2)$ for the second half of the time step (part (c) in Algorithm II).

*2. 5. Simulation details*

All the electronic structure calculations were carried out based on the DMM method (a linear scaling DFT approach) using the CONQUEST code [38–41]. In the DMM method, the elements in the matrix $L$ are nonzero only when the distance between the centres of the support functions is less than a chosen cutoff distance; throughout the calculations, we used a cutoff of 16 bohr. We employed the PBE exchange-correlation functional [43] and used norm-conserving pseudopotentials [44] to treat valence–core interactions. In the extended

Lagrangian BOMD simulations, we chose the dissipation parameter $K = 5$ and the corresponding values for the parameters $\kappa$, $\alpha$, and $\{c_k\}$ reported by Niklasson *et al.* [12]. In the Nosé–Hoover chain algorithm, a chain of 5 thermostats was coupled to the ionic degrees of freedom and the 15th-order Yoshida–Suzuki integrator was used in the Nosé–Hoover chain part of the evolution operator. The thermostat frequency was set to 500 cm$^{-1}$.

To check the stability and feasibility of our integration scheme, we performed several NVT extended Lagrangian BOMD simulations. First, we performed simulations on a bulk silicon system at 300, 600, and 900 K. A time step of 1 fs was used and the total time length of the simulations was 50 ps. We used a 64-atom silicon model in a cubic cell of length 10.86 Å (the density $\rho = 2.33$ g cm$^{-3}$) with periodic boundary conditions. A tolerance of $\varepsilon = 10^{-5}$ was used in the DMM procedure. A single-$\zeta$ basis set was used with an energy grid cutoff of 216 Ry. Second, using the same computational conditions as described above, we also performed a simulation on a 3072-atom system at 300 K for 12 ps. Third, we performed a simulation on a cubic silicon carbide (SiC) system consisting of 64 atoms at a higher temperature (1500 K). A time step of 1 fs was used and the time length of the simulations was 10 ps. We used a cubic cell of length 8.716 Å (the density $\rho = 3.22$ g cm$^{-3}$) with periodic boundary conditions. A tolerance of $\varepsilon = 10^{-5}$ was used in the DMM procedure. A single-$\zeta$ basis set was used with an energy grid cutoff of 187 Ry.

**Algorithm I.** Integration scheme that combines the extended Lagrangian BOMD formulation and the Nosé–Hoover chain algorithm. See the text for details of the symbols.

---

a) Propagate the Nosé–Hoover chain part for $\frac{\Delta t}{2}$: $U_{\text{NHC}}(\frac{\Delta t}{2}) = \prod_{j=1}^{n_{\text{ys}}} \exp(iL_{\text{NHC}}w_j\frac{\Delta t}{2})$

---

b) Propagate the ionic and electronic degrees of freedom for $\Delta t$:

$\dot{\mathbf{R}}_I \leftarrow \dot{\mathbf{R}}_I + \frac{\Delta t}{2M_I}\mathbf{F}_I$,

$\mathbf{R}_I \leftarrow \mathbf{R}_I + \Delta t \dot{\mathbf{R}}_I$,

$P(t+\Delta t) = 2P(t) - P(t-\Delta t) + \kappa(D(t) - P(t)) + \alpha \sum_{k=0}^{K} c_k P(t - k\Delta t)$,
$D(t+\Delta t) = \text{SCF}[P(t+\Delta t)]$,
Update the forces: $\mathbf{F}_I = -\partial E_{\text{tot}}/\partial \mathbf{R}_I$,

$\dot{\mathbf{R}}_I \leftarrow \dot{\mathbf{R}}_I + \frac{\Delta t}{2M_I}\mathbf{F}_I$.

---

c) Propagate the Nosé–Hoover chain part for $\frac{\Delta t}{2}$: $U_{\text{NHC}}(\frac{\Delta t}{2}) = \prod_{j=1}^{n_{\text{ys}}} \exp(iL_{\text{NHC}}w_j\frac{\Delta t}{2})$

---

**Algorithm II.** Integration scheme that combines the extended Lagrangian BOMD formulation using the DMM method and the Nosé–Hoover chain algorithm. See the text for details of the symbols.

---

a) Propagate the Nosé–Hoover chain part for $\frac{\Delta t}{2}$: $U_{\text{NHC}}(\frac{\Delta t}{2}) = \prod_{j=1}^{n_{\text{ys}}} \exp(iL_{\text{NHC}}w_j\frac{\Delta t}{2})$

---

b) Propagate the ionic and electronic degrees of freedom for $\Delta t$:

$\dot{\mathbf{R}}_I \leftarrow \dot{\mathbf{R}}_I + \frac{\Delta t}{2M_I}\mathbf{F}_I$,

$\mathbf{R}_I \leftarrow \mathbf{R}_I + \Delta t \dot{\mathbf{R}}_I$,

$X(t+\Delta t) = 2X(t) - X(t-\Delta t) + \kappa(LS(t) - X(t)) + \alpha \sum_{k=0}^{K} c_k X(t - k\Delta t)$,
Compute initial $S$ and $S^{-1}$,
Density matrix minimization using $XS^{-1}$ as an initial guess
Update the forces: $\mathbf{F}_I = -\partial E_{\text{tot}}/\partial \mathbf{R}_I$,

$\dot{\mathbf{R}}_I \leftarrow \dot{\mathbf{R}}_I + \frac{\Delta t}{2M_I}\mathbf{F}_I$.

---

c) Propagate the Nosé–Hoover chain part for $\frac{\Delta t}{2}$: $U_{\text{NHC}}(\frac{\Delta t}{2}) = \prod_{j=1}^{n_{\text{ys}}} \exp(iL_{\text{NHC}}w_j\frac{\Delta t}{2})$

---

## 3. Results and discussion

To begin with, we compared the energy conservation for both conventional BOMD and extended Lagrangian BOMD. In both approaches, the temperature was well controlled by the Nosé–Hoover chain thermostat (figure 1(a)); however, the behaviour of the energy conservation differed significantly between the two methods. When the DMM method was combined with the conventional BOMD method, the extended energy (the sum of the physical and the thermostat energies) decreased sharply within a few picoseconds (figure 1(b)), which indicates a significant deviation from the ground-state phase space. Even though we increased the tighter tolerance for the DMM procedure (from $\varepsilon = 10^{-5}$ to $10^{-6}$), a rapid, systematic energy drift was still observed because of the incompleteness of the convergence, which in practice would not be achieved [10, 11, 31].

In the extended Lagrangian BOMD method, on the other hand, the conserved quantity remained stable with no systematic energy drift (figure 1(b)), in agreement with previous work [12–20, 31]. The number of iterations in the DMM procedure was moderate (the maximum number of iterations for $\varepsilon = 10^{-5}$ was 7 cycles per time step). This energy conservation, in spite of the additional complexity in the propagation for both the electronic and ionic degrees of freedom, supports the applicability and robustness of the extended Lagrangian BOMD scheme for a wide range of electronic structure calculation methods.

The stark difference between the two approaches can be explained by the time-reversal symmetry in the propagation of the electronic degrees of freedom [10, 11, 31]. In the conventional BOMD approach, the matrix $L$ obtained from the previous time step was used for an initial guess for the DMM method. Although this approach is efficient in the DMM procedure (one iteration was needed for $\varepsilon = 10^{-5}$; three iterations for $\varepsilon = 10^{-6}$), it breaks the time-reversal symmetry in the propagation of the electronic degrees of freedom. This broken time-reversibility of the electronic degrees of freedom behaves like a heat sink for the physical system, albeit controlled by the thermostatting algorithm.

In the conventional BOMD formalism, the potential energy rapidly decreased owing to the broken time symmetry (figure 1(c)); it resulted in a rapid decrease in the forces acting on atoms (not shown), which prohibits physically meaningful MD simulations. In the extended Lagrangian BOMD formalism, the potential energy kept physically relevant values (figure 1(c)). This is because the initial value used for the density matrix is obtained in a time-reversible way (See equation (19)).

To assess the feasibility and stability of our integration scheme, we performed NVT simulations on bulk silicon for 50 ps at three different temperatures: 300, 600, and 900 K. The time evolution of the conserved quantity and the potential energy for the simulations is shown in figure 2. To evaluate the energy fluctuations, we calculated the mean absolute deviations of the conserved quantity: $1.03 \times 10^{-5}$, $2.01 \times 10^{-5}$, and $3.19 \times 10^{-5}$ a.u. per atom for 300, 600, and 900 K, respectively. These show that the conserved quantity for these simulations was stable with acceptable fluctuations, with the magnitude of the fluctuations increasing with temperature (figure 2). In line with the energy conservation, neither systematic energy drift in the potential energy nor systematic decrease in the forces was observed (not shown), which indicates the stability of our integration scheme.

Next, we investigated the distribution of temperatures obtained from the extended Lagrangian BOMD simulations in the canonical ensemble (figure 2(a)). In addition to the simulations on bulk silicon, we also performed an NVT-MD simulation on a silicon carbide (SiC) system consisting of 64 atoms at 1500 K. This simulation is considered to be a more severe test because the auxiliary electronic degrees of freedom can be indirectly affected by elevated thermal fluctuations. The average temperatures in the extended Lagrangian BOMD simulations were 300.5, 599.6, 898.5, and 1507.1 K, respectively. Thus, the average temperatures (the first moment) were correctly obtained within the relative error of 0.5% with respect to target temperatures. We also checked the variance of the temperatures. In theory, the variance (the second moment) for temperature is given by

$$\langle (\delta T)^2 \rangle = \frac{2\langle T \rangle^2}{3N}$$

(20)

for a three-dimensional system of $N$ atoms [31]. The temperature fluctuations obtained from the MD trajectories were in good agreement with the above relation (figure 2(b)), which validates the fidelity of our integration scheme.

To evaluate the scope and limitations of our integration scheme, we investigated the effects of two key parameters: the time step for MD simulations and the tolerance value in the DMM method. We performed extended Lagrangian BOMD simulations on a bulk silicon system at 900 K using different time steps: 1.0, 2.5, 3.0, and 3.5 fs. The conserved quantity remained stable for 50 ps when time steps of 1.0 and 2.5 fs were used (for larger time steps, the quantity was not well preserved). Second, we checked the effects of the $L$ tolerance value, by performing simulations on a silicon carbide system at 1500 K for 10 ps. The mean absolute deviations of the conserved quantity for $\varepsilon = 10^{-5}$ and $10^{-6}$ were $2.74\times10^{-5}$ and $8.04\times10^{-6}$ a.u. per atom, respectively; clearly, the conserved quantity fluctuates less with $\varepsilon = 10^{-6}$ than with $\varepsilon = 10^{-5}$, though both are sufficiently stable for practical simulations. When a time step of 0.5 fs was used in combination with a tolerance of $\varepsilon = 10^{-6}$, the mean absolute deviation of the conserved quantity was $1.74\times10^{-5}$ a.u. per atom. Our analyses imply that the stability of the integration may be managed by adjusting modest tolerance values in the DMM procedure with appropriate time steps.

To validate our integration scheme in terms of system size, we performed an NVT run (12 ps) on a crystalline system containing 3072 silicon atoms. We found that the stability of the conserved quantity remained similar to the one obtained from a 64-atom system (figure 4). The fluctuation of the potential energy per atom was smaller, since the number of atoms was increased from 64 to 3072 atoms. We checked that the number of electrons was well controlled for this system (the electron number $N_e = 12288 \pm 1.3\times10^{-4}$). The maximum number of iterations in the DMM procedure was 6 cycles per time step. The stability of our integration scheme in terms of system size is not unexpected because $\mathcal{O}(N)$ calculations on millions of atoms can be performed [29]; therefore, we expect the stability of our integration scheme presented here to be maintained for systems containing more than thousands of atoms.

## 4. Conclusion

We have provided an integration scheme for canonical-ensemble extended Lagrangian BOMD in the context of the Liouville operator formulation, where the Nosé–Hoover chain thermostat was used as a temperature-control method. In addition, we developed a molecular dynamics scheme for NVT extended Lagrangian BOMD scheme in conjunction with the DMM method in the linear scaling DFT CONQUEST code. To verify the stability of our integration scheme, we performed NVT extended Lagrangian BOMD simulations on a 64-atom Si system at 300, 600, and 900 K and on a 64-atom SiC system at 1500 K. The results show long-term stability of the energy conservation. Our analysis shows that the average temperatures and the variances in the canonical ensemble were correctly obtained. Furthermore, we demonstrated an NVT run of a system containing more than 3000 Si atoms: the stability of the conserved quantity remained similar to the one obtained from the 64-atom silicon system. The present study not only provide useful information on a stable integration scheme for canonical-ensemble extended Lagrangian BOMD simulations, but also suggests the applicability and robustness of the extended Lagrangian BOMD scheme for a wide range of electronic structure calculations including $\mathcal{O}(N)$ methods.

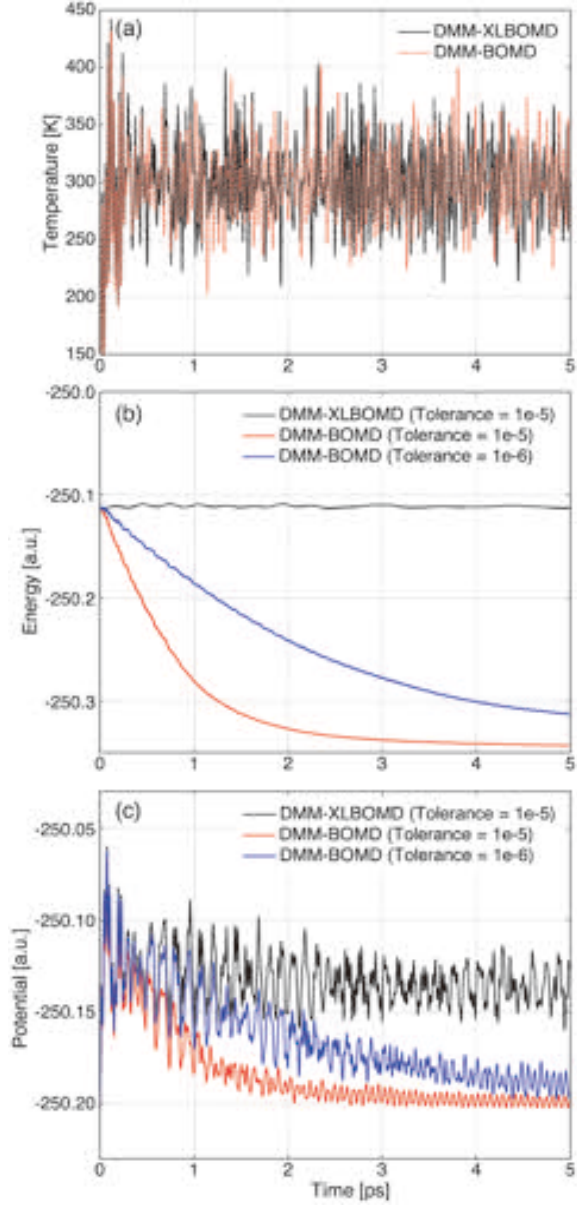

**Figure 1.** Comparison between conventional and extended Lagrangian $\mathcal{O}(N)$ BOMD simulations (crystalline silicon) in the canonical ensemble. (a) Time evolution of temperature during conventional (dashed red) and extended Lagrangian (black) BOMD simulations. (b) Time evolution of the extended energy during conventional (red: $\varepsilon = 10^{-5}$; blue $\varepsilon = 10^{-6}$) and extended Lagrangian (black) BOMD simulations. (c) Time evolution of the potential energy during conventional (red: $\varepsilon = 10^{-5}$; blue: $\varepsilon = 10^{-6}$) and extended Lagrangian (black) BOMD simulations.

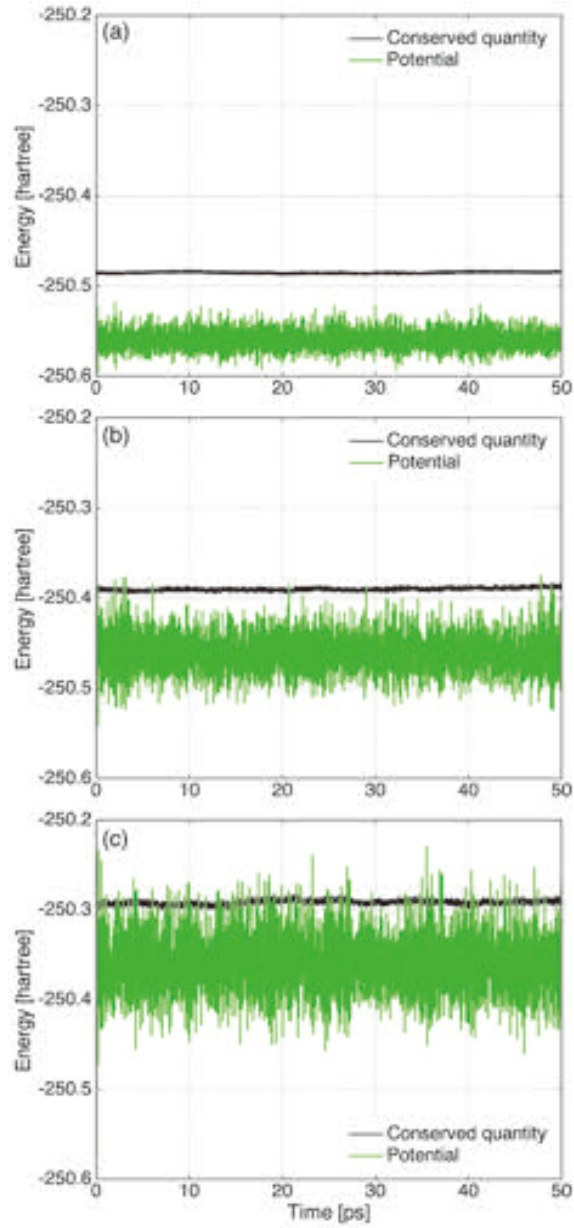

**Figure 2.** Time evolution of the conserved quantity (black) and the potential energy (green) obtained from NVT simulations on bulk silicon at 300 K (a), 600 K (b), and 900 K (c) in the framework of the extended Lagrangian BOMD formalism in combination with the DMM method.

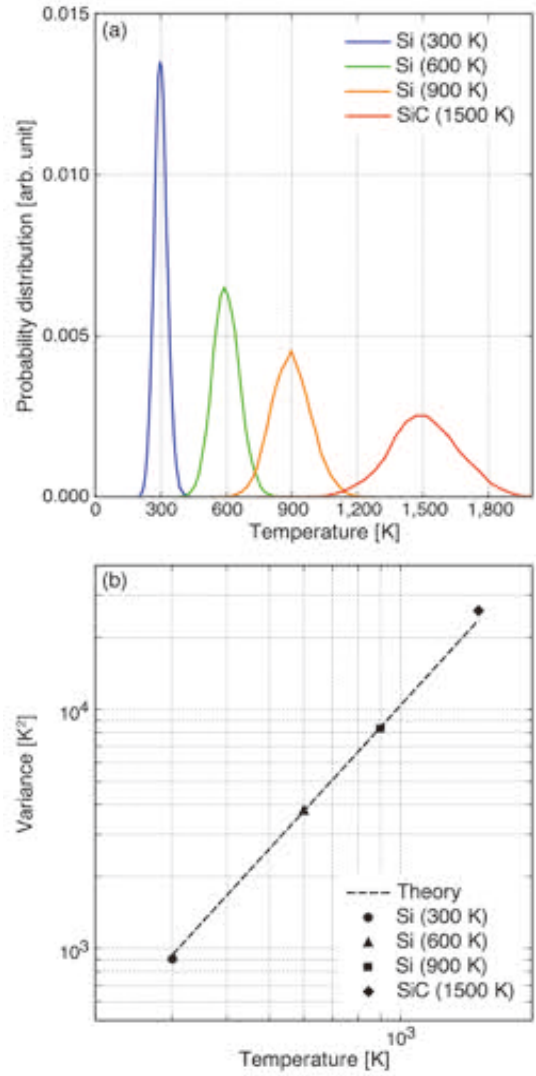

**Figure 3.** (a) Distributions of temperature $T$ obtained from simulations on bulk Si at 300 K (blue), 600 K (green), and 900 K (orange) and a simulation on cubic SiC at 1500 K (red) in the framework of extended Lagrangian BOMD in combination with the DMM method. (b) Second moments (variances) as a function of temperature. Circle: Si at 300 K; triangle: Si at 600 K; square: Si at 900 K; diamond: SiC at 1500K. The dashed line indicates the theoretical value given by $\langle(\delta T)^2\rangle = 2\langle T\rangle^2/3N$, where $N$ is the number of atoms.

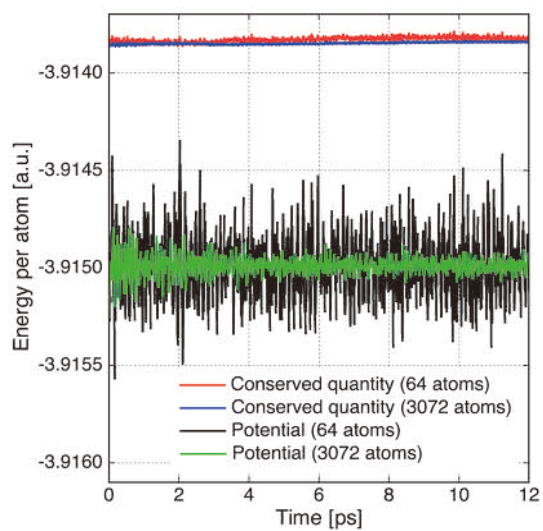

**Figure 4.** Time evolution of the conserved energy per atom for a 64-atom system (red) and a 3072-atom system (blue) and the potential energy per atom for a 64-atom system (black) and a 3072-atom system (green). In both NVT simulations, the target temperature in the canonical ensemble was set to 300 K.

Acknowledgments

We thank Prof Yoshitada Morikawa for suggesting the implementation of the Nosé–Hoover chain thermostat in CONQUEST and Dr Takao Otsuka for discussing the conservation of the energy. TS thanks Dr Masato Sumita for his encouragement during the early days of this work; Dr Toshiaki Iitaka for helpful discussions and his critical review of the manuscript; and Ms Iriko Kaneko for her technical help in producing the graphics in this paper. TH thanks Dr Takashi Ikeda for discussing the implementation of the Nosé–Hoover chain thermostat. This work was supported in part by "World Premier International Research Center Initiative on Materials Nanoarchitectonics (WPI-MANA)" and "Exploratory Challenge on Post-K computer" (Frontiers of Basic Science: Challenging the Limits) by the Ministry of Education, Culture, Sports, Science, and Technology (MEXT), Japan. TH thanks "Interactive Materials Science Cadet Program," funded by the Program for Leading Graduate Schools, MEXT, Japan. Part of our extended Lagrangian BOMD simulations was performed on Numerical Materials Simulator at NIMS.

References

[1] Hohenberg P and Kohn W 1964 *Phys. Rev.* **136** B864

[2] Kohn W and Sham L J 1965 *Phys. Rev.* **140** A1133

[3] Yu H S, Li L S and Truhlar D G 2016 *J. Chem. Phys.* **145** 130901

[4] Car R and Parrinello M 1985 *Phys. Rev. Lett.* **55** 2471

[5] Marx D and Hutter J 2000 *Modern Methods and Algorithms of Quantum Chemistry*, 2nd ed., edited by J. Grotendorst (Jülich: John von Neumann Institute for Computing)

[6] Pulay P and Fogarasi G 2004 *Chem. Phys. Lett.* **386** 272

[7] Herbert J M and Head-Gordon M 2005 *Phys. Chem. Chem. Phys.* **7** 3269

[8] VandeVondele J, Krack M, Mohamed F, Parrinello M, Chassaing T and Hutter J, 2005 *Comput. Phys. Commun.* **167** 103

[9] Kuhne T D, Krank M, Mohamed F R and Parrinello M 2007 *Phys. Rev. Lett.* **98** 066401

[10] Niklasson A M N, Tymczak C J and Challacombe M 2006 *Phys. Rev. Lett.* **97** 123001

[11] Niklasson A M N 2008 *Phys. Rev. Lett.* **100** 123004

[12] Niklasson A M N, Steneteg P, Odell A, Bock N, Challacombe M, Tymczak C J, Holmström E, Zheng G and Weber V 2009 *J. Chem. Phys.* **130** 214109

[13] Cawkwell M J, Coe J D, Yadav S K, Liu X Y and Niklasson A M N 2015 *J. Chem. Theory Comput.* **11** 2697

[14] Aradi B, Niklasson A M N and Frauenheim T 2015 *J. Chem. Theory Comput.* **11** 3357

[15] Souvatzis P and Niklasson A M N 2013 *J. Chem. Phys.* **139** 214102

[16] Steneteg P, Abrikosov I A, Weber V and Niklasson A M N 2010 *Phys. Rev.* B **82** 075110

[17] Cawkwell M J and Niklasson A M N 2012 *J. Chem. Phys.* **137** 134105

[18] Arita M, Bowler D R and Miyazaki T 2014 *J. Chem. Theory Comput.* **10** 5419

[19] Mniszewski S M, Cawkwell M J, Wall M E, Mohd-Yusof J, Bock N, Germann T C and Niklasson A M N 2015 *J. Chem. Theory Comput.* **11** 4644

[20] Negre C F A, Mniszewski S M, Cawkwell M J, Bock N, Wall M E and Niklasson A M N 2016 *J. Chem. Theory Comput.* **12** 3063

[21] Goedecker S 1999 *Rev. Mod. Phys.* **71** 1085

[22] Bowler D R and Miyazaki T 2012 *Rep. Prog. Phys.* **75** 036503

[23] Khaliullin R Z, VandeVondele J and Hutter J, 2013 *J. Chem. Theory Comput.* **9** 4421

[24] Niklasson A M N, Mniszewski S M, Negre C F A, Cawkwell M J, Swart P J, Mohd-Yusof J, Germann T C, Wall M E, Bock N, Rubensson E H and Djidjev H 2016 *J. Chem. Phys.* **144** 234101

[25] Ju S and Liang X J 2012 *J. Appl. Phys.* **112** 064305

[26] France-Lanord A, Blandre E, Albaret T, Merabia S, Lacroix D and Termentzidis K 2014 *J. Phys.: Condens. Matter* **26** 055011

[27] Drewitt J W E, Jahn S, Sanloup C, de Grouchy C, Garbarino G and Hennet L 2015 *J. Phys.: Condens. Matter* **27** 105103

[28] Karplus M and McCammon J A 2002 *Nat. Struct. Mol. Biol.* **9** 646

[29] Bowler D R and Miyazaki T 2010 *J. Phys.: Condens. Matter* **22** 074207

[30] VandeVondele J, Borštnik U and Hutter J 2012 *J. Chem. Theory Comput.* **8** 3565

[31] Martínez E, Cawkwell M J, Voter A F and Niklasson A M N 2015 *J. Chem. Phys.* **142** 154120

[32] Nosé S 1984 *J. Chem. Phys.* **81** 511

[33] Hoover W G 1985 *Phys. Rev.* A **31** 1695

[34] Martyna G J, Klein M L and Tuckerman M 1992 *J. Chem. Phys.* **97** 2635

[35] Martyna G J, Tuckerman M E, Tobias D J and Klein M L 1996 *Mol. Phys.* **87** 1117

[36] Frenkel D and Smit B 2002 *Understanding Molecular Simulation: From Algorithms to Applications*, 2nd ed. (San Diego: Academic Press) ch 4, pp 74–76

[37] Tuckerman M E and Parrinello M 1994 *J. Chem. Phys.* **101** 1302

[38] Hernández E, Gillan M J and Goringe C M, 1996 *Phys. Rev.* B **53** 7147

[39] Bowler D R, Miyazaki T and Gillan M J, 2001 *Comput. Phys. Commun.* **137** 255

[40] Bowler D R, Choudhury R, Gillan M and Miyazaki T 2006 *Phys. Status Solidi* B **243** 989

[41] Otsuka T, Miyazaki T, Ohno T, Bowler D R and Gillan M J 2008 *J. Phys.: Condens. Matter* **20** 294201

[42] Press W H, Teukolsky S A, Vetterling W T and Flannery B P 1992 *Numerical Recipes in Fortran 77: The Art of Scientific Computing,* 2nd ed. (Cambridge: Cambridge University Press) ch 2, pp 49–50

[43] Perdew J P, Burke K and Ernzerhof M, 1996 *Phys. Rev. Lett.* **77** 3865

[44] Troullier N and Martins J L 1991 *Phys. Rev.* B **43** 1993